\documentclass[aps,twocolumn,prl,superscriptaddress,showpacs,preprintnumbers,amsmath,amssymb]{revtex4}

\usepackage{graphicx}
\usepackage{color}

\begin{document}

\title{Full spin switch  effect for the superconducting
current in a superconductor/ferromagnet thin film heterostructure}

\author{P. V. Leksin}
\affiliation{Zavoisky Physical-Technical Institute, Kazan
Scientific Center of Russian Academy of Sciences, 420029 Kazan,
Russia}
\author{N. N. Garif'yanov}
\affiliation{Zavoisky Physical-Technical Institute, Kazan
Scientific Center of Russian Academy of Sciences, 420029 Kazan,
Russia}
\author{I. A. Garifullin}
\email[]{ilgiz_garifullin@yahoo.com} \affiliation{Zavoisky
Physical-Technical Institute, Kazan Scientific Center of Russian
Academy of Sciences, 420029 Kazan, Russia}
\author{J. Schumann}
\affiliation{Leibniz Institute for Solid State and Materials
Research IFW Dresden, D-01171 Dresden, Germany}
\author{H.~Vinzelberg}
\affiliation{Leibniz Institute for Solid State and Materials
Research IFW Dresden, D-01171 Dresden, Germany}
\author{V. Kataev}
\affiliation{Leibniz Institute for Solid State and Materials
Research IFW Dresden, D-01171 Dresden, Germany}
\author{R. Klingeler}
\affiliation{Leibniz Institute for Solid State and Materials
Research IFW Dresden, D-01171 Dresden, Germany}
\author{O. G. Schmidt}
\affiliation{Leibniz Institute for Solid State and Materials
Research IFW Dresden, D-01171 Dresden, Germany}
\author{B. B\"{u}chner}
\affiliation{Leibniz Institute for Solid State and Materials
Research IFW Dresden, D-01171 Dresden, Germany}

\date{\today}

\begin{abstract}
Superconductor/ferromagnet (S/F) proximity effect theory predicts
that the superconducting critical temperature of the F1/F2/S or
F1/S/F2 trilayers for the parallel orientation of the F1 and F2
magnetizations is smaller than for the antiparallel one. This
suggests a possibility of a controlled switching between the
superconducting and normal states in the S layer. Here, using the
spin switch design F1/F2/S theoretically proposed by Oh {\it et
al.} [Appl. Phys. Lett. {\bf 71}, 2376 (1997)], that comprises a
ferromagnetic bilayer separated by a non-magnetic metallic spacer
layer as a ferromagnetic component, and an ordinary superconductor
as the second interface component, we have successfully realized a
full spin switch effect for the superconducting current.



\end{abstract}

\maketitle

The antagonism of superconductivity (S) and ferromagnetism (F)
consists of strong suppression of superconductivity by
ferromagnetism because ferromagnetism requires parallel (P) and
superconductivity requires antiparallel (AP) orientation of spins.
The exchange splitting of the conduction band in strong
ferromagnets which tends to align electron spins parallel is
larger by orders of magnitude than the coupling energy for the AP
alignment of the electron spins in the Cooper pairs in
conventional superconductors. Therefore the singlet pairs with AP
spins of electrons will be destroyed by the exchange field. For
this reason the Cooper pairs can penetrate into an F-layer only
over a small distance. The characteristic distance of decay of the
pairing function in the F-layer is $\xi_F=(4\hbar D_F/I)^{1/2}$,
where $D_F$ and $I$ are the diffusion coefficient and the exchange
splitting of the conduction band in the F-layer, respectively
\cite{Radovic}. For pure Fe the value of $\xi_F$ is less than 1 nm
(see, e.g., \cite{Lazar}).

The physical origin of the spin switch effect based on the S/F
proximity effect  relies on the idea to control the pair-breaking,
and hence the superconducting transition temperature $T_c$, by
manipulating the mutual orientation of the magnetizations of the
F-layers in a heterostructure comprising, e.g., two F- and one
S-layer in a certain combination. This is because the mean
exchange field from two F-layers acting on Cooper pairs in the
S-layer is smaller for the AP orientation of the magnetizations of
these F-layers compared to the P case. Historically, the first
paper devoted to the realization of the spin switch effect by
manipulating the mutual orientation of the magnetizations of the
F-layers has been published by Deutscher and Meunier in 1969
\cite{Deutscher}. They studied FeNi/In/Ni trilayer and obtained a
surprisingly large difference in $T_c$ between the AP and P
orientations of the magnetizations. The reason for this effect has
not been clarified up to now. Clinton and Johnson \cite{Clinton}
have developed a superconducting valve which uses the magnetic
fringe fields at the edges of the F film of a $\mu$m size. These
fringe fields can be varied in magnitude by changing the mutual
orientation of the magnetization of two F layers separated by a
nonmagnetic (N) spacer layer. In their F/N/F/S construction a
dielectric interlayer between the F/N/F valve and the bridge in
the S-layer with a width of 1 $\mu$m  was formed artificially.
Thus, a direct contact between magnetic and superconducting parts
of the sample was absent similar to the case studied in Ref.
\cite{Deutscher}. The possibility to develop a switch based on the
S/F proximity effect has been theoretically substantiated in 1997
by Oh {\it et al.} \cite{Oh}. They proposed the F1/N/F2/S layer
scheme where an S-film is deposited on top of two F-layers, F1 and
F2, separated by a thin metallic N-layer. The thickness of F2
should be smaller than $\xi_F$ to allow the superconducting pair
wave function to penetrate into the N-layer. Two years later a
different construction based on a trilayer F/S/F thin film
structure was proposed theoretically by Tagirov \cite{Tagirov} and
Buzdin {\it et al.} \cite{Buzdin,Buzdin1}. Several experimental
works confirmed the predicted influence of the mutual orientation
of the magnetizations in the F/S/F structure on $T_c$ (see, e.g.,
\cite{Gu,Potenza,Moraru,Miao}). However, the difference in $T_c$
between the AP and P orientations $\Delta T_c$ turns out to be
smaller than the width of the superconducting transition $\delta
T_c$ itself. Hence a full switching between the normal and the
superconducting state was not achieved. Implementation of a design
similar to the F1/N/F2/S layer scheme by Oh {\it et al.} \cite{Oh}
with a [Fe/V]$_n$ antiferromagnetically coupled superlattice
instead of a single F1/N/F2 trilayer \cite{Westerholt,Westerholt1}
can not be switched from the AP to P orientations of the
magnetizations instantaneously. At the same time the analysis of
the temperature dependence of the critical field has shown that
implicitly $\Delta T_c$ of this system can be as large as 200 mK
at $\delta T_c \sim$100 mK.

Comparison of the results obtained for both proposed constructions
of the spin switches gives grounds to suppose that the scheme by
Oh {\it et al.} may be the most promising for the realization of
the full spin switch effect for the superconducting current in an
S/F thin film heterostructure. In order to get a maximum spin
switch effect we concentrated our efforts on the optimization of
the materials' choice and of the specific geometry of the
F1/N/F2/S scheme by Oh {\it et al.} \cite{Oh} and have fabricated
a set of samples which show a full switching between the
superconducting and the normal state when changing the mutual
orientation of the magnetizations of F1 and F2 layers.

For the layer sequence AFM/F1/N/F2/S to be deposited on the single
crystalline MgO substrate (Fig.~1) we have chosen the following
set of materials: cobalt oxide for the antiferromagnetic (AFM)
layer that plays a role of the bias layer which pins the
magnetization of the F1 layer; Fe for the ferromagnetic F1- and
F2-layers; Cu as a normal metallic N-layer; and finally In for the
S-layer. While the Fe/Cu/Fe trilayer as a ferromagnetic component
looks ordinary, the choice of the S-component in our scheme is of
primary importance. Considering that only few superconducting
materials do not form a solid solution or intermetallic compounds
with iron (they are Pb and In) and therefore do not form the
otherwise unavoidable intermixed interface region
\cite{Garifullin} we decided to concentrate on indium.
\begin{figure}[t]
\centering{\includegraphics[width=0.3\columnwidth,angle=-90,clip]{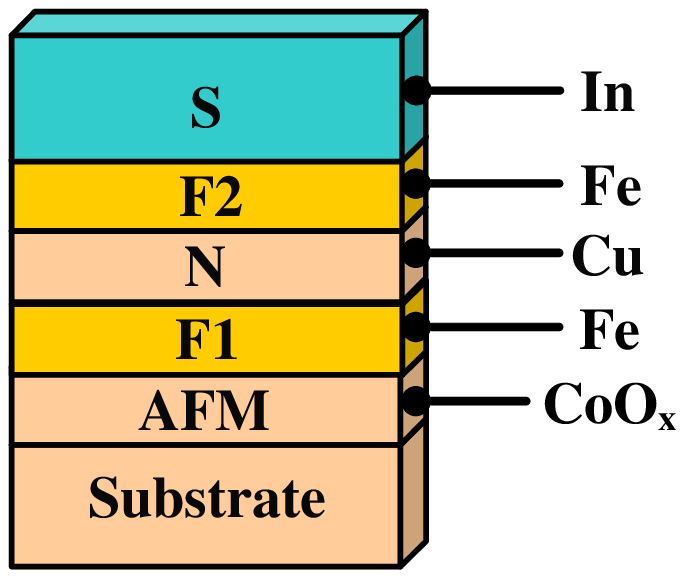}}
\caption{(Color online) Design of the studied samples.}
\end{figure}
The sample preparation was done by electron beam evaporation in an
ultra-high vacuum chamber within a closed vacuum cycle. The base
pressure was $2\cdot 10^{-8}$ mbar. All deposition experiments
were carried out on room-temperature substrates. The thickness of
the growing films was measured by a quartz crystal monitor system.
The Co oxide films were prepared by a two-step process consisting
of the evaporation of a metallic Co film followed by the plasma
oxidation converting Co into CoO$_x$ layer.

The residual resistivity ratio $RRR$=$R$(300K)/$R$(4K) for all
studied samples is similarly high for all studied samples (see
Table 1) evidencing a high purity of the deposited In layers.

The indium film in our samples is a type I superconductor with
parallel and perpendicular critical fields $H_c^{\parallel} \sim
220$ Oe and $H_c^{\perp} \sim 20$ Oe, respectively, at $T$=2 K. In
view of this anisotropy we have taken care to avoid the appearance
of the perpendicular component of an external field larger than 2
Oe. This means that we adjusted the sample plane position with an
accuracy better than 2 degrees relative to the direction of the dc
external field. The easy axis of the magnetization which is
induced by residual magnetic fields in our vacuum system was
directed parallel to the long axis of the sample. The parameters
of the studied samples are shown in Table 1. Along with the spin
switch samples \#\# 3 -- 5 we prepared for control purposes an
indium thin film sample (\#1) and a reference sample comprising an
indium layer and only one F-layer (\#2R).
\begin{table}
\caption{Experimental parameters of the studied samples}
\begin{center}
\begin{tabular}{|c|c|c|c|c|c|c|c|c|}
\hline
 & \multicolumn{5}{c|}{Layers' thickness (nm)} &  & \\
 \cline{2-6} \raisebox{1.5ex}[0cm][0cm]
 {Sample} & $\rm CoO_x$ & Fe & Cu & Fe & In & $RRR$&
  $\delta T_c$ (mK) & $\Delta T_c$
  (mK) \\
 \hline
       1 &  &  &  &  & 220 & 43 & 7 & 0$\pm 2$ \\
\hline
    2R &  &  &  & 0.5 & 230 &35 & 15 & 0$\pm 3$ \\
\hline
3 & 4 & 2.4 & 4 & 0.5 & 230 & 47& 7 & 19$\pm 2$ \\
\hline
4 & 4 & 2.9 & 4 & 0.6 & 230 & 41 & 13 & 12$\pm 3$  \\
\hline
5 & 4 & 2.6 & 4 & 2.6 & 230 & 44& 50 & -2$\pm 8$ \\
\hline
\end{tabular}
\label{tc_tab}
\end{center}
\end{table}

In our work we put emphasis on the experimental determination of
both the hysteresis magnetization behavior and the current
in-plane transport measurements, enabling a correlation between
both properties.

In a first step the in-plane magnetic hysteresis loops of sample
\#3 in the direction of the magnetic field along the easy axis
were measured by a SQUID magnetometer and is shown in Fig. 2a.
This step is necessary to find out the Fe-layers' magnetization
behavior and to determine the magnetic field range where AP and P
states can be achieved. The sample was cooled down in a magnetic
field of +4 kOe applied parallel to the sample plane and measured
at $T = 4$\,K. The magnetic field was varied from +4 kOe to - 6
kOe and back again to the value of +4 kOe. Both limits correspond
to the orientation of the magnetizations of the F1- and F2-layers
parallel to the applied field. For the studied sample by
decreasing the field from +4 kOe to the field value of the order
of +50 Oe the magnetization of the free F2 layer starts to
decrease. At the same time the magnetization of the F1-layer is
kept by the bias CoO$_x$ layer until the magnetic field of -4 kOe
is reached. Thus, in the field range between -0.3 and -3.5 kOe the
mutual orientation of two F-layers is antiparallel. Below $H$=
-3.5 kOe the magnetization of the F1 layer starts to change it's
value and at the field of the order of -4.5 kOe magnetizations of
both Fe layers become parallel. This corresponds to a further
step-like decrease of the total magnetization. Qualitatively
similar hysteresis loops were obtained for samples \#4 and 5. The
minor hysteresis loops on the low field scale obtained with
decreasing the field from +4 kOe down to -1 kOe and increasing it
again up to +1 kOe are shown in Figs. 2b, 2c and 2d for samples
\#3, 4 and 5, respectively. It is necessary to note that later
when performing the transport measurements, the magnetic field did
not reach values beyond $\pm$1 kOe.\\
\begin{figure}[t]
\centering{\includegraphics[width=1.1\columnwidth,angle=-90,clip]{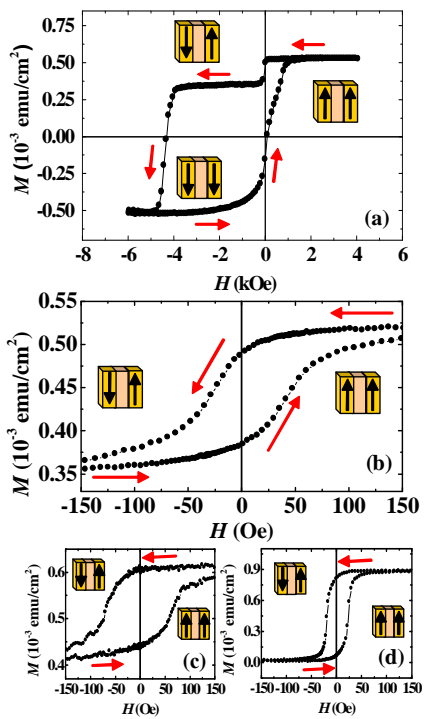}}
\caption{ (Color online) (a): Magnetic hysteresis loop for sample
\#3. Panels (b), (c) and (d) show parts of the minor hysteresis
loops for samples \#3, 4 and 5, respectively, obtained by
decreasing the magnetic field from +4 kOe down to -1 kOe and
increasing it up to +1 kOe. The amplitude of the minor hysteresis
loops is proportional to the thickness of the free F2 layer.
Coercive and saturation field is smallest for sample \#5.}
\end{figure}
For the transport study we used another system which also enables
a very accurate control of the real magnetic field acting on the
sample. This field was generated by a high homogeneous
electromagnet. The magnetic field value was measured with an
accuracy of $\pm$0.3 Oe using a Hall probe. The temperature of the
sample was monitored by the 230 $\Omega$ Allen-Bradley resister
thermometer which is particular sensitive in the temperature range
of interest. Therefore the accuracy of the temperature control
within the same measurement cycle below 2 K was better than $\pm
2\div 3$ mK.
\begin{figure}
\centering{\includegraphics[width=1.0\columnwidth,angle=-90,clip]{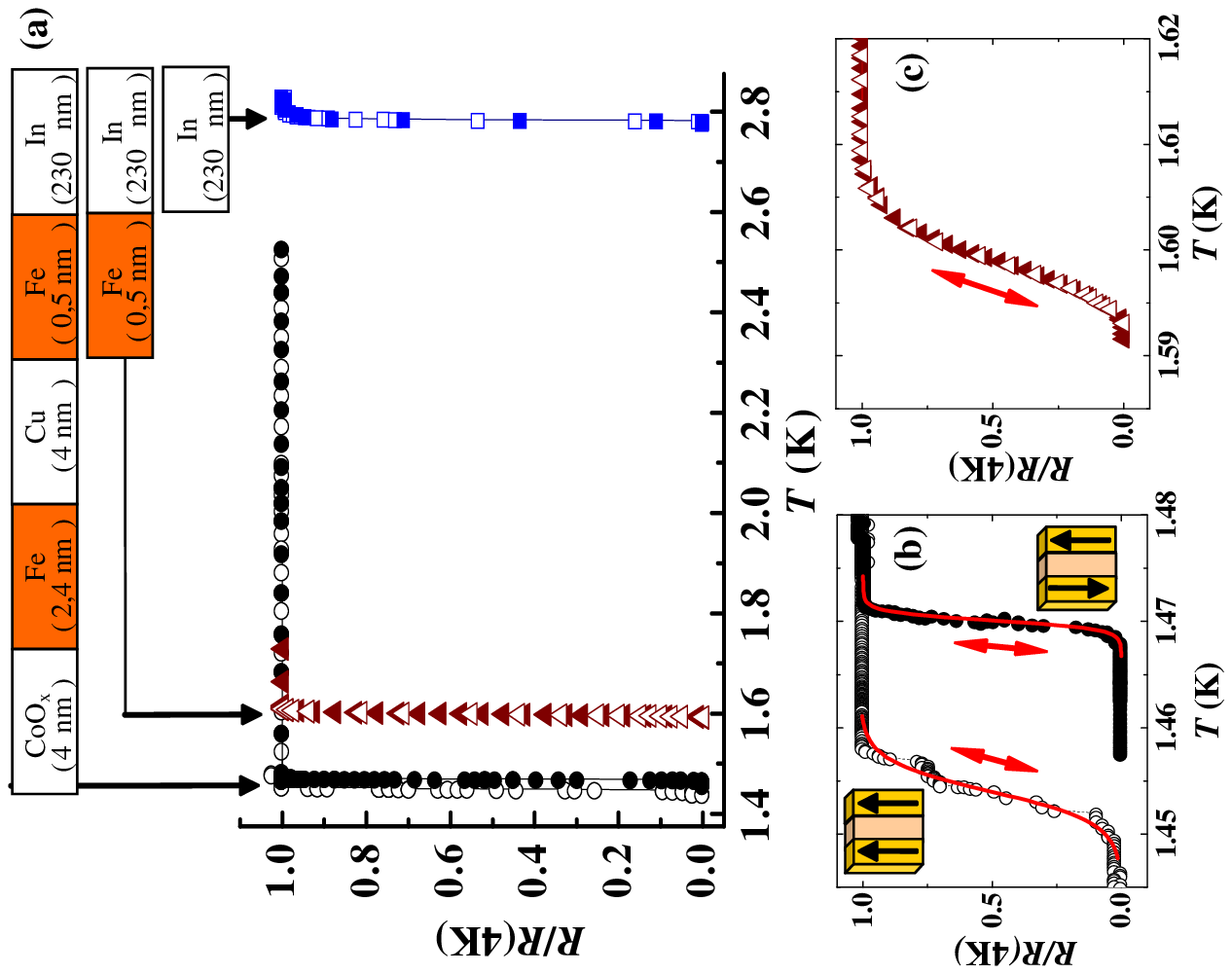}}
\caption{(Color online) (a): Overview of the resistivity
transition curves. The spin valve sample \#3 is shown by open
($H_0$= +110 Oe) and closed ($H_0$= -110 Oe) circles. For the
reference sample \#2R the data are depicted by open ($H_0$=+110
Oe) and closed ($H_0$=-110 Oe) triangles. For the pure In sample
the data are presented by open ($H_0$=+110 Oe) and closed
($H_0$=-110 Oe) squares. The superconducting transition
temperature for the reference sample \#2R is lower than for the
single In layer and higher than for the sample \#3. (b) and (c)
demonstrate the details of the superconducting transitions for
sample \#3 and the reference sample \#2R, respectively.}
\end{figure}
In order to study the influence of the mutual orientation of the
magnetizations on $T_c$ we have cooled the samples down from room
to a low temperature at a magnetic field of 4 kOe applied along
the easy axis of the sample just as we did it when performing the
SQUID magnetization measurements. For this field both F-layers'
magnetizations are aligned parallel (see the magnetic hysteresis
loops in Fig. 2). Then at the in-plane magnetic field value of
$H_0 = \pm$ 110\,Oe the temperature dependence of the resistivity
$R$ was recorded. In the following we focus on the spin valve
sample \# 3 (see Fig. 3). For this sample the difference in $T_c$
for different magnetic field directions is clearly seen (see Fig.
3b with an enlarged temperature scale). The superconducting
transition temperature for the AP orientation of the
magnetizations occurs at a temperature exceeding the $T_c$ for the
P orientation of the sample by 19~mK. We also performed similar
resistivity measurements of the reference sample \#2R with only
one Fe layer (see Table 1). For this sample we found $T_c$=1.60 K,
which does not depend on the magnetic field direction (see Fig.
3c). This $T_c$ value is lower than that for the In single layer
film (sample \#1) and higher than for sample \#3 (Fig. 3). This
means that $T_c$ is suppressed by the F2 layer and in turn is
sensitive to the influence of the F1 layer separated from the
superconducting In layer by a 0.5 nm thick F2 Fe layer and 4 nm
thick Cu layer. As can be expected from the the S/F proximity
theory, with increasing the thickness of the free F2 layer $\Delta
T_c$ decreases and becomes practically zero for the  2.6 nm thick
F2 layer (see Table 1).

The observed shift of the superconducting transition temperature
$\Delta T_c$=19\,mK is not the largest one among the data
published before (cf., e.g., $\Delta T_c \simeq 41$ mK at $\delta
T_c \sim $100 mK in Ref. \cite{Moraru}). However, it is very
important is that $\delta T_c$  is significantly {\it larger} than
the superconducting transition width $\delta T_c$ which is as
small as $\sim$7 mK for sample \#3 at $H_0$=110\,Oe. This opens a
possibility to switch off and on the superconducting current
flowing through our samples {\it completely} within the
temperature range corresponding to the $T_c$-shift by changing the
mutual orientation of magnetization of F1 and F2 layers. To
demonstrate this we have performed the measurements of the
resistivity of sample \#3 by sweeping slowly the temperature
within the $\Delta T_c$ interval and switching the magnetic field
between +110 and -110\,Oe. This central result of our study is
shown in Fig. 4. It gives straightforward evidence for a complete
on/off switching of the superconducting current flowing through
the sample.  To the best of our knowledge, this is the first ever
example of the realization of a full spin switch design for a
superconducting current in F/S structures with a perfect contact
at each interface.
\begin{figure}
\centering{\includegraphics[width=0.5\columnwidth,angle=-90,clip]{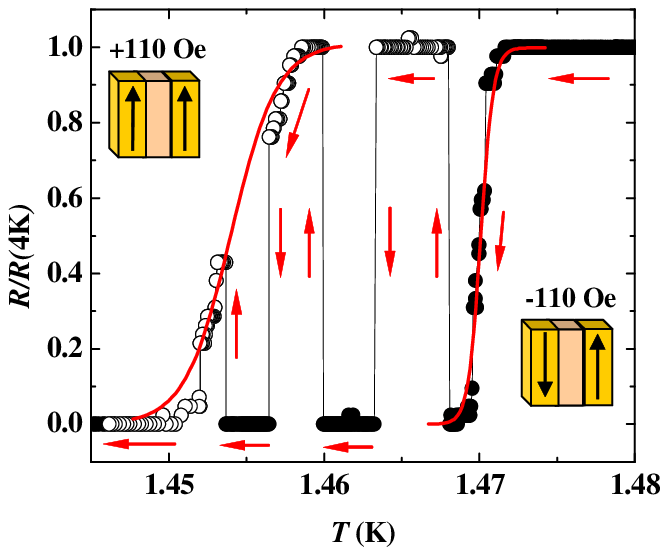}}
\caption{(Color online) Switching between normal and
superconducting states in the spin valve sample \#3 during a slow
temperature sweep by applying the magnetic field $H_0$= -110 Oe
(closed circles) and $H_0$=+110 Oe (opened circles) in the sample
plane. Solid lines correspond to the superconducting transition
curves measured resistively (cf. Fig. 3).}
\end{figure}

For sample \#3 (as well as for sample \#4) the obtained result is
in agreement with the expectation based on the S/F proximity
effect theory, namely that $T_c$ for the AP orientation of
magnetizations is higher than $T_c$ for the P orientation. In the
end it is not surprising, since all necessary prerequisites to
realize the theoretical idea of Oh {\it et al.}~\cite{Oh} are
fulfilled in this sample: (i) -the thickness of the F2-layer is
smaller than the coherence length $\xi_F$; (ii) -owing to the
absence of the intermixed region an atomically sharp highly
transparent metallic interface between the S- and the F-layer has
been achieved. The combination of (i) and (ii) makes a penetration
of the Cooper pair wave function into the F2-layer and further
into the N-layer possible. Finally, the high quality of the iron
layers yields magnetization hysteresis curves with sharp well
defined steps enabling a well controlled switching of the mutual
orientation of the magnetization of the F-layers by application of
relatively small magnetic fields. This is essential for the
control of the decay of the Cooper pair wave function.

In summary, we have presented the first, to the best of our
knowledge, experimental realization of the spin switch for the
superconducting current in that we have achieved a complete
switching on and off a superconducting current flowing through a
sample by changing the mutual orientation of magnetizations of the
F layers. Our results provide a compelling experimental
confirmation of theoretical predictions by Oh {\it et al.}
\cite{Oh} for the spin switch device thereby suggesting the S/F
proximity effect as the operation principle of our switch.

\begin{acknowledgements}
The authors are grateful to Professors  A. V. Volkov, L. R.
Tagirov and Ya. V. Fominov for useful discussions. The work was
supported by the RFBR (grants No. 08-02-00098, 08-02-91952-NNIO)
and the German-Russian cooperation project of the DFG (grant 436
RUS 113/936/0-1).
\end{acknowledgements}


\begin{thebibliography} {99}

\bibitem{Radovic} Z. Radovi\'{c}, L. Dobrosavljevi\'{c}-Gruji\'{c}, A.
I. Buzdin, J. R. Clemm. Phys.~Rev. B {\bf 38},  2388 (1988).

\bibitem{Lazar} L. Lazar, K. Westerholt, H. Zabel, L. R. Tagirov, Yu. V. Goryunov,
N. N. Garif'yanov and I.A. Garifullin. Phys.~Rev. B {\bf 61}, 3711
(2000).

\bibitem{Deutscher} G. Deutscher and F. Meunier. Phys.~Rev.~Lett. {\bf 22}, 395
(1969).

\bibitem{Clinton} T. W. Clinton and M. Johnson. Appl.~Phys.~Lett. {\bf 70},
1170 (1997).

\bibitem{Oh} S. Oh, D. Youm, and  M. R. Beasley.  Appl. Phys. Lett.
{\bf 71}, 2376 (1997).

\bibitem{Tagirov} L. R. Tagirov. Phys. Rev. Lett. {\bf 83},
2058 (1999).

\bibitem{Buzdin} A. I.  Buzdin, A. V. Vedyaev. and N. N.
Ryzhanova. Europhys. Lett. {\bf 48}, 686 (1999).

\bibitem{Buzdin1} I. Baladi\'{e}, A. Buzdin, N. Ryzhanova, and A. Vedyayev. Phys.
Rev. B {\bf 64}, 054518 (2001).

\bibitem{Gu} J. Y. Gu, C. Y. You,  J. S. Jiang, J. Pearson, Ya. B. Bazaliy
and S. D. Bader.  Phys. Rev. Lett. {\bf 89}, 267001 (1-4) (2002).

\bibitem{Potenza} A. Potenza and C. H. Marrows. Phys. Rev. B {\bf 71},
180503(R) (1-4) (2005).

\bibitem{Moraru} I. C. Moraru , Jr. W. P. Pratt  and N. O. Birge.
Phys. Rev. Lett. {\bf 96}, 037004 (1-4) (2006); Phys. Rev. B {\bf
74}, 220507(R) (1-4) (2006).

\bibitem{Miao} Guo-Xing Miao, Ana V. Ramos, and Jagadeesh Moodera.
Phys.~Rev.~Lett. {\bf 101}, 137001 (2008).

\bibitem{Westerholt} K. Westerholt, D. Sprungmann, H. Zabel, R. Brucas,
B. Hj\"{o}rvarsson, D. A. Tikhonov. and I. A. Garifullin. Phys.
Rev. Lett. {\bf 95}, 097003 (1-4) (2005).

\bibitem{Westerholt1} G. Nowak, H. Zabel, K. Westerholt, I. Garifullin,
M. Marcellini, A. Liebig. and Hj\"{o}rvarsson. Phys. Rev. B {\bf
78}, 134520 (1-12) (2008).

\bibitem{Garifullin} I. A. Garifullin, N. N. Garif'yanov, R. I. Salikhov.
Izv. RAN: Ser. Fiz. {\bf 71}, 280 (2007) [Bul. RAN: Phys. {\bf
71}, 272 (2007)].

\end{thebibliography}
\end{document}